\documentclass[10pt,paper,twocolumn]{IEEEtran}
\usepackage[T1]{fontenc}
\usepackage{graphicx}
\usepackage{caption2}
\usepackage{array}
\usepackage{makecell,colortbl,xcolor,booktabs,multirow}
\usepackage{setspace,subfigure}
\usepackage{cite}
\usepackage{hhline}
\usepackage{diagbox}
\usepackage{url}
\ifCLASSINFOpdf
\else
\fi
\usepackage{amsmath}
\usepackage[cmintegrals]{newtxmath}
\begin{document}

\title{\LARGE Distributed Swarm Learning for Internet of Things at the Edge: \\Where Artificial Intelligence Meets Biological Intelligence}

\author{Yue~Wang,~\IEEEmembership{Senior Member,~IEEE}, Zhi~Tian,~\IEEEmembership{Fellow,~IEEE}, Xin~Fan,~\IEEEmembership{Member,~IEEE}, \\ Yan~Huo,~\IEEEmembership{Senior Member,~IEEE}, Cameron~Nowzari,~\IEEEmembership{Member,~IEEE},
and Kai~Zeng,~\IEEEmembership{Member,~IEEE}
\IEEEcompsocitemizethanks{
\IEEEcompsocthanksitem This work was partly supported by the US NSF grants 1939553, 2003211, 2128596, 2136202 and 2231209, and the Virginia Research Investment Fund CCI grant 223996.
}}

\maketitle

\section*{Abstract}
With the proliferation of versatile Internet of Things (IoT) services, smart IoT devices are increasingly deployed at the edge of wireless networks to perform collaborative machine learning tasks using locally collected data, giving rise to the edge learning paradigm. Due to device restrictions and resource constraints, edge learning among massive IoT devices faces major technical challenges caused by the communication bottleneck, data and device heterogeneity, non-convex optimization, privacy and security concerns, and dynamic environments. To overcome these challenges, this article studies a new framework of distributed swarm learning (DSL) through a holistic integration of artificial intelligence and biological swarm intelligence. Leveraging efficient and robust signal processing and communication techniques, DSL contributes to novel tools for learning and optimization tailored for real-time operations of large-scale IoT in edge wireless environments, which will benefit a wide range of edge IoT applications.

\IEEEpeerreviewmaketitle

\section{Introduction}
\noindent
Smart Internet of Things (IoT)
are becoming the workhorse at wireless edge, where
a tremendous amount of valuable data directly collected from the edge environments together with the resurgence of machine learning (ML) stimulate the latest trend of artificial intelligence (AI) at the edge, a.k.a. edge learning. However, conventional  ML methods are not suitable for edge learning, since they hinge on collecting raw data from local devices and thus raising privacy leakage and security risk. Alternatively, federated learning (FL) has emerged to allow distributed learning while keeping data locally~\cite{mcmahan2017communication}, which has led to fruitful attempts in implementing AI applications among  distributed workers such as personal smart phones.

The success of FL typically relies on ideal learning settings and wireless environments, where smart phones possess powerful computing capability 
and communicate over secure networks with stable connectivity.
But, these assumptions become invalid in many practical edge IoT applications where low-cost IoT devices are equipped with limited communication and computation capability.
When neural networks are employed, the size of model parameters goes large, posing a major challenge in transmitting the local model updates in the FL training stage. Stochastic gradient descent (SGD) is widely applied for model training in FL, where independent and identically distributed (i.i.d.) data samples are assumed at local workers to ensure unbiased estimates and good empirical performance. This is however not the case in edge IoT, as IoT devices only access small-volume data.
All these factors give rise to the following challenges to IoT-driven edge learning.

{\bf \em Communication bottleneck}. The large size of learning models and huge number of distributed workers incur high communication costs for edge learning in IoT networks, while low-cost IoT devices are prone to power constraints, bandwidth limits, communication errors, and link failures.

{\bf \em Data and device heterogeneity}.
Edge learning faces both device heterogeneity on IoT hardware capability and link quality,
and statistical heterogeneity of local training data at different workers, a.k.a., the non-i.i.d. data issue, which may dramatically
degrade the  performance of edge learning.

{\bf \em Non-convex optimization}.
Gradient-based algorithms are subject to local optimum traps in solving non-convex problems, such as when training neural networks with nonlinear activations. This issue is aggravated in distributed settings, especially when IoT workers only collect small-volume data.

{\bf \em Privacy and security concerns}.
Standard FL performs well in idealized attack-free network settings but is vulnerable to outliers, eavesdroppers, privacy breach, and Byzantine attacks, which inevitably exist in real IoT systems and edge networks.

{\bf \em Dynamic environments}.
In some real-time edge IoT applications, IoT workers have to deal with streaming data while the dynamics underlying the data and the surrounding  environments are unknown and time-varying, where traditional techniques designed for the static case fail to work effectively.

Although some of these challenges have been recently investigated in the literature of FL for IoT~\cite{khan2021federated}, main efforts are found on modification and customization of standard FL. Most results stem from the AI perspective alone, but neglect unique characteristics of edge IoT, including the large population of IoT devices, limited capability of each IoT device, and  local streaming data of small volume.
Intriguingly, biological organisms in nature having swarm intelligence, even if individually weak, have successfully demonstrated superior strength in collectively searching the optimal solutions, recovering from errors, and adapting to environment changes. Since these attributes are sought out by IoT-driven edge learning, biological intelligence (BI) is expected to boost efficiency and robustness of edge learning among massive low-cost IoT.

 \begin{figure}
  	\centering 		
  \includegraphics[width=3.56in]{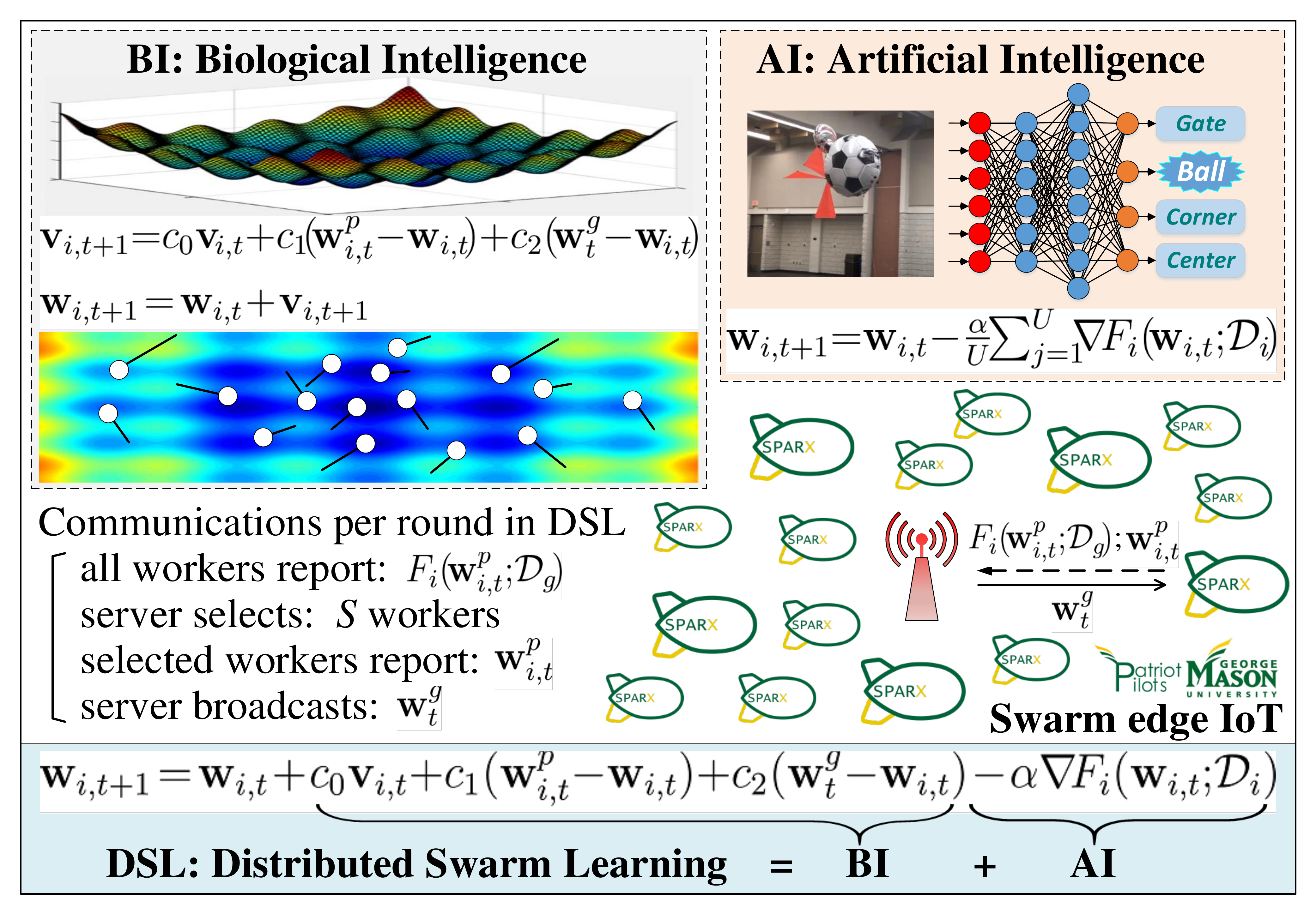}
\caption{\small Distributed swarm learning (DSL) for edge IoT. For each worker $i$, its local model parameter vector $\mathbf{w}_{i,t}$ is updated through local velocity $\mathbf{v}_{i,t}$ and gradient $\nabla F_i$ terms from PSO and FL respectively, where $\mathbf{w}_{i,t}^p$ records its own historical best variable, $\mathbf{w}_{t}^g$ is the globally best variable of the swarm,  $F_i(\cdot)$ is the local loss evaluated on the local dataset $\mathcal{D}_i$, and $c_0$, $c_1$, and $c_2$ are weights to balance exploration and exploitation. Upon updating $\mathbf{w}_{i,t}$ and subsequently the locally best variable $\mathbf{w}_{i,t}^p$, each worker calculates its fair-value loss $F_i(\mathbf{w}_{i,t}^p; \mathcal{D}_g)$ using a small global dataset $\mathcal{D}_g$ that is introduced and made available to all workers (Section IV.A). When a central server is present, it collects these scalar loss values to perform adaptive worker selection, so that only one or a few workers with the lowest losses are selected to transmit their locally best variables $\mathbf{w}_{i,t}^p$ to the server for updating the global $\mathbf{w}_{t}^g$ via over-the-air mean aggregation (Section III.A). The idea of DSL can also be extended and applied for the fully decentralized topology, where each worker communicates with its neighbors and makes autonomous decisions (Section VII).}\vspace{0.08in}
\label{fig:DSL_EdgeIoT}
\end{figure}

To bridge this gap, this article investigates a holistic integration of AI and BI, by leveraging swarm intelligence and cooperative gains among massive IoT workers to develop efficient, robust and adaptive distributed learning techniques. A new edge learning paradigm, called distributed swarm learning (DSL)~\cite{JSTSP2022DSL}, is proposed by connecting the AI-enabled FL with the BI-inspired particle swarm optimization~\cite{kennedy1995particle}, as illustrated in Figure \ref{fig:DSL_EdgeIoT}.
The rest of this article systematically studies DSL from the following perspectives: design of its conceptual framework, development of its enabling techniques, performance evaluation, and discussion on future directions.

\section{Framework of Distributed Swarm Learning}
\noindent
Consider an IoT-driven edge learning paradigm, where massive low-cost IoT workers collaboratively learn the parameters of a common model by minimizing a task-oriented loss function, given limited communication, computation, and data at each worker. Further, the learning performance is impacted by bandwidth limits, malicious attacks, node malfunctions, link failures, and heterogeneity in devices and data.

\subsection{Federated Learning}
\noindent
Standard FL is originally designed in ideal learning and network settings \cite{mcmahan2017communication}, without the additional constraints on IoT devices and edge environments.
Therein the minimization of the loss function is typically carried out by the SGD algorithm, whose performance highly relies on i.i.d. data samples and error-free transmissions. Local workers individually update local models, and then the global model or averaged gradient is calculated from all local updates, which works as the initial point for local updating at the next iteration. Computation and communication take place at all local workers in every iteration, until FL convergence. For non-convex problems, gradient-based FL solutions may converge to undesired local optima and there is lack of mechanisms to escape these traps.

\subsection{Particle Swarm Optimization}
\noindent
Particle swarm optimization (PSO), as a BI-inspired algorithm, can solve complicated optimization problems without assumption on convexity \cite{kennedy1995particle}. Mimicking swarm behavior in animal flocks, PSO runs based on the movement of particles and the collaboration of swarms to search for an optimal solution.
The position of each particle presents a possible solution to the problem, and its velocity denotes the updating direction for the next step. To find the globally optimal solution in swarm, particles collaborate in updating velocities and positions in an iterative manner. Notably, the velocity is updated as a weighted combination of three sub-directions: inertia of the previous updating direction, individual direction towards each particle's own historical best variable, and social direction towards the globally best variable found by the entire swarm. Such a weighted combination serves a mechanism for exploration-exploitation tradeoffs in swarm optimization. Specifically, the inertia weight is decreasing over iterations to tune the solution search process from exploration to exploitation. The other two weights are random variables 
to indicate the random exploration level at individual particles and the exploitation level in swarm, respectively.

\subsection{Distributed Swarm Learning}
\noindent
In the model updating process, the velocity in PSO and the gradient in FL play a similar role as the updating direction, but their updating principles are different. In PSO, the velocity is a weighted combination of three sub-directions, where the collaboration is reflected in the third socially updated sub-direction towards the up-to-date globally best variable. As a gradient-free stochastic optimization method, PSO is good at collaboratively searching for the global optimal solutions to complex problems thanks to the built-in exploration-exploitation mechanism and the swarm nature, but its convergence is typically slow. Moreover, 
PSO assumes a globally common loss function for all workers in order to enable collaboration, which no longer holds in distributed learning problems characterized by data-dependent local loss functions. In contrast, FL is a gradient-based learning algorithm with fast convergence. However, it is subject to local optimum traps and suffers from high communication costs in massive IoT systems when all workers transmit their local updates.

The aforementioned pros and cons of FL and PSO motivate to bridge distributed learning with swarm optimization techniques to make the best use of both artificial and biological intelligence. There are few recent attempts of applying PSO ideas to improve FL performance. In~\cite{qolomany2020particle}, FL is used for learning, while PSO is simply applied to search the optimal hyperparameters. In~\cite{park2021fedpso}, PSO and FL are combined in a simplistic manner for the idealized distributed settings with i.i.d. data and no attacks. The work~\cite{park2021fedpso} builds on an implicit assumption that a common loss function is available to all local workers, which trivializes the assessment of the globally best model through single-worker selection. However, in distributed learning problems, the loss function is only {partially observable} at local workers, which is data-dependent and hence different across workers. Thus, the method in~\cite{park2021fedpso} fails to work properly in edge IoT with non-i.i.d. data at local workers.

To fill the identified technical gaps above,
DSL is recently proposed as
a novel distributed learning framework tailored to edge IoT~\cite{JSTSP2022DSL}, 
for which new theoretical foundations, algorithm designs, and analytical approaches are developed.
%
The DSL framework is built on three essential components. 1) {\em Holistic integration of AI and BI}: The fundamental concept and basic ideas of the DSL framework are mathematically illustrated in Figure~\ref{fig:DSL_EdgeIoT}, where the local models are updated from both AI and BI perspectives through a combination of the gradient-based SGD term and the gradient-free PSO terms, to balance performance and convergence speed. 2) {\em Employing a global dataset}: To facilitate worker collaboration, the DSL generates a very small amount of globally shared dataset and use it to assess the local loss of all workers, which helps to accurately identify the per-worker best variable and select a few best workers for model variable aggregation.  3) {\em Efficient and robust communication and aggregation}: The DSL principle is implemented efficiently by several key techniques including multi-worker selection and over-the-air analog aggregation for global variable updating.
These key DSL operations will be elaborated in ensuing sections. At the outset, DSL provides the following major benefits.

  {\bf \em Rapid convergence}.
  In DSL, the velocity in PSO is combined with the gradient in SGD, which amounts to embedding a BI-inspired exploration-exploitation mechanism to SGD methods. Thus, DSL algorithms are expected to expedite the convergence, which outperforms either PSO or SGD alone.

  {\bf \em Reduced communication overheads}.
  Since only a small number of local workers are selected  to share their local model updates for collaboration in the swarm, the communication costs and transmit power consumed by DSL can be saved dramatically with parsimonious transmission strategies.

  {\bf \em Resistance to local optimum traps}.
  The velocity updating step in DSL employs random weights and exploration-exploitation mechanisms, which results in an increase chance for the solutions to jump to new positions, and hence
facilitates the swarm intelligence to escape from local optimum traps.

  {\bf \em Relaxed device restrictions}.
  By using over-the-air analog aggregation, multiple workers
  simultaneously transmit their local model variables over the same time-frequency resources.
  Hence, device restrictions
  in communication bandwidth and resource constraints minimally affect the DSL operations.

  {\bf \em Robustness to node/link failures}.
  Per the updating rule of DSL, even if some  workers do not respond to the request for sharing local updates, the rest workers are still able to work as long as the network connectivity is not broken, which enhances the robustness of DSL against node and/or link failures.

\section{Efficient Distributed Swarm Learning}
\noindent
A body of literature has been devoted to address the communication efficiency of FL, via gradient sparsification, message quantization, and infrequent transmissions of local model updates. They either compress the information to be transmitted or drop less-informative transmissions prior to aggregation of local updates. Recently, a promising technique emerges as FL over-the-air \cite{fan2021joint}, which exploits the fact that the model-aggregation operation in FL matches the waveform-superposition property of wireless analog multi-access channels (MAC). It enables concurrent transmissions from all distributed workers and its analog aggregation over-the-air reduces the bandwidth consumption. All these methods still encourage all participating workers to exchange  local updates, which are not tailored to edge IoT systems and result in tremendous communication costs among massive IoT devices. To save device energy consumption and reduce total communication cost in edge networks, DSL incorporates parsimonious transmission and aggregation schemes with the built-in exploration-exploitation mechanism in swarm.

\subsection{Efficient Communication and Computation}
\noindent
Recall the different collaboration mechanisms between FL and PSO: In FL, all workers report their local updates for global averaging; while in PSO, only one worker is selected as the up-to-date global best to share its local update with others. For the purpose of the best worker selection in PSO, each worker is also requested to share the minimum value of its historical local loss functions to compare with that of others.
FL and PSO work as the two extreme cases under the DSL framework. To collect the benefits from both sides, the communication and aggregation protocols in DSL need to be carefully designed with theoretical backing.

\subsubsection{Communication Censoring}
As highlighted in Figure 1, all the workers need to report their individual historical best local function values, based on which a few best workers with the lowest losses are selected for the calculation of the global model variables in each iteration.
Even though the loss function values are scalar, such a reporting process still consumes large communication overhead as the total number of workers goes large in edge IoT. Note that the reported values are sorted for worker selection, while the sorting operation accommodates some level of value approximations. It prompts to introduce a communication-censoring strategy at each worker to save individual communication cost. That is, each worker assesses how much its current local best function value differs from its previously recorded one, and reports only when the difference is large enough to exceed a censoring threshold. Otherwise, the worker does not report to save transmission, and worker selection is made based on its previously reported value as an approximate of its current value. The censoring process is autonomous based on the judiciously designed local censoring threshold to guarantee convergence~\cite{xu2021coke}.

\subsubsection{Adaptive Multi-Worker Selection}
After collecting or approximating workers' historical best local function values through the communication censoring scheme, it is essential to deploy a multi-worker selection strategy by deciding $S$, the number of selected workers who will contribute their local historical best variables to the global variable updating. Note that DSL encourages local workers to move along the social direction towards the global best variable as a means of exploitation of swarm collaboration. To effectively fulfill the exploration-exploitation mechanism with communication efficiency, an adaptive strategy can be employed to increase the value of $S$ along the iterations. Intuitively, a smaller $S$ at the early stage of iterations allows individual workers to focus on exploring the solutions locally at little communication overhead, and then a larger $S$ during later iterations brings more workers to contribute to the collective swarm intelligence for high model training accuracy at convergence.

\subsubsection{Over-The-Air Analog Aggregation}
In DSL, the global variable updating only requires the averaged updates of $S$ selected workers rather than their individual local variables.
{DSL over-the-air} is a direct and efficient way of implementing global variable updating,
which employs
analog aggregation based transmission to enable the $S$ selected workers to simultaneously transmit their local variables over the same time-frequency resources, as long as the aggregated waveform represents the averaged updates after proper transmit power control\cite{fan2021joint}. Thus, the total consumed bandwidth is minimized and independent of $S$. Since $S$ reflects the degrees of freedom in DSL to control the exploration-exploitation mechanism in swarm intelligence, adaptive multi-worker selection strategy can be co-designed and assessed based on the impact of the over-the-air transmission on the learning performance. In addition, DSL can further incorporate other efficiency-enhancing strategies such as 1-bit compressed sensing,
which amounts to combining compression, quantization and concurrent transmission to attain impressive efficiency \cite{fan20211}.

\subsection{Convergence Behavior}
\noindent
DSL offers superiority in solving non-convex  problems, which mainly benefits from the inherent capability in escaping from local optimum traps owning to the random weights and exploration-exploitation mechanisms embedded in PSO. Such a benefit of swarm intelligence are intuitive and widely appreciated. From the theoretical side, the global convergence properties of PSO have been of interest to the computational science community~\cite{Clerc2002particle}. While the superiority of PSO lies in its high probability of achieving global convergence, PSO itself does not necessarily improve the order of convergence speed, unless joint optimization for parameter selection and resource allocation is carried out at the system level. In convergence analysis of DSL, the metric of interest is the expected convergence rate of local workers, which is used to evaluate the convergence of stochastic algorithms. A closed-form expression for the expected convergence rate achieved by DSL is derived as a function of the number of communication rounds $T$ and the PSO-related exploration-exploitation parameters in~\cite{JSTSP2022DSL}, which is bounded on the order of $\mathcal{O}({1}/{T})$.

\subsection{Joint Optimization of Communication and Computation}
\noindent
The convergence behavior of DSL unveils a fundamental connection between edge communications and distributed learning, which provides a fresh perspective to measure how the parameter choice of wireless systems and the hyperparameter design of computational algorithms affect the performance of edge learning. Guided by the theoretical results from convergence analysis, joint optimization of learning parameter determination, worker selection, and transmit power control is formulated as minimizing the loss function given limited transmit power and bandwidth.
It amounts to a network resource optimization problem that minimizes the learning error subject to the maximum power constraints of low-cost IoT devices~\cite{fan2021joint, fan20211},
which yields the jointly optimized worker selection and power control variables for DSL operations.

\section{Robust Distributed Swarm Learning}
\noindent
In edge IoT networks, the local data samples of small volume typically turn out to be non-i.i.d. across workers. Further, there exist malfunctioning workers and even malicious attackers. In addition, link failure contributes to unreliable transmissions. All these issues call for robust measures for DSL.

\subsection{Global Dataset Generation}
\noindent
As a vital component of the DSL framework, a globally shared dataset is introduced to play dual roles:  to provide fair-value scores of local models for multi-worker selection and to alleviate the non-i.i.d. issue. Centralized data sharing methods would collect raw data samples from individual workers to form global datasets \cite{tuor2021overcoming}, but raise privacy concerns.  
To keep raw data private at local workers, robust data augmentation methods are developed to form a global dataset, based on federated generative adversarial network (FedGAN) by training a neural generator in a distributed manner~\cite{Li2022IFL}.
This generator can be used to generate synthetic data samples as the globally shared dataset for DSL. Numerical results show that a very small amount of global data (say $1\%$ of all data) is adequate to deliver the expected high learning performance~\cite{JSTSP2022DSL}. The global data can be shared prior to DSL operations, and the required resources in sharing and local storage are quite low.

\subsection{Robust Measures against Non-i.i.d. Issues}
\noindent
Having generated a global dataset via the FedGAN-based data augmentation, the global dataset is further divided into two parts: one for model training to alleviate the  non-i.i.d. issue, and the other for scoring worker quality against Byzantine attacks, respectively. To alleviate the non-i.i.d. issue, a globally shared training dataset is distributed among all local workers. This simple idea turns out to be effective, as evaluated by a statistical analysis approach that is developed to evaluate the impact of model weight divergence on the learning performance of DSL~\cite{JSTSP2022DSL}. 
The model weight divergence is related to the distance enlargement between the non-i.i.d. data distributions of local workers and the overall data population distribution. The non-i.i.d. issue can be further tackled by employing a regularized total variation in the local loss function.
Such a regularized penalty term enforces local variables to be close to each other by reducing the distance from current local updates
to the previous global variables,
which alleviates the impact from non-i.i.d. datasets.

\subsection{Robust Measures against Byzantine Attacks}
\noindent
The rest global dataset is used to score the local loss function values, so that the training data and scoring data are decoupled and i.i.d. Meanwhile, it also serves to screen and kick out potential Byzantine attackers who want to cheat in local scoring by reporting confusing local model updates. Analog aggregation transmission is known to enhance the privacy thanks to the inherent inaccessibility to individual local updates.
While DSL over-the-air closes the doors to model inversion attacks, it leaves the windows open for adversaries to perform Byzantine attacks. The work \cite{fan2022bev} indicates that even a single Byzantine fault may destroy distributed learning via analog aggregation. The impacts of Byzantine attacks to DSL are studied under the swarm setting at the edge, taking into account of the integrated nature of AI and BI in DSL~\cite{JSTSP2022DSL}. Based on the theoretical analysis~\cite{fan2022bev}, a best effort voting (BEV) based transmission power control policy can be applied to against Byzantine attacks for DSL over-the-air. 

\subsection{Robust Measures against Node/Link Failures}
\noindent
Thanks to the analog aggregation transmission, the worker selection is not subject to constrained communication bandwidth, which allows to select more workers to collectively update the global model variables at no extra bandwidth. Leveraging such multi-worker diversity, DSL can enhance the resiliency to unreliable transmission links and malfunctioning workers. Further, to cope with wireless fading, truncated channel inversion method is developed for FL over-the-air~\cite{zhu2019broadband}. When some workers experience deep fading channels, the truncated channel inversion method is employed to weed out those workers with low channel gains. For DSL over-the-air, the truncated channel inversion policy needs to be designed with attention to the incorporation of the BI terms. By jointly determining the truncation thresholds with the multi-worker selection scheme in DSL over-the-air, the truncation rules can effectively balance both the communication and computation aspects of distributed learning over wireless fading channels.

\section{Online Adaptive Distributed Swarm Learning}
\noindent
In real-time IoT applications, the data is received as a streaming sequence and local workers need to process the data adaptively in an online fashion. Further, the dynamics underlying the streaming data can be varying, and the function of interest itself may vary over time with unknown dynamics. In this sense, the original learning problems become online learning (OL) problems, where dynamic optimization is involved.

The performance of OL algorithms is expressed and evaluated in terms of regrets, including static regret and dynamic regret. The sublinear static regret bound can be achieved by using an online gradient descent based algorithm or an online alternating direction method via multipliers (ADMM) \cite{Liu2021Dqc}. When the loss function changes over time, dynamic regret is a more suitable performance metric where the OL algorithm minimizes the loss function per iteration. The standard OL approaches face limitations in dynamic edge networks, which rely on a super-node that not only processes network-wide streaming data information but also assesses the dynamic loss functions of all local workers over iterations. However, in large-scale IoT networks, it is impossible to centrally process all such information needed for OL algorithms, nor can it access the dynamic loss functions. In this case, distribute workers with partial information need to cooperate in solving a global OL problem. To this end, DSL offers a natural solution since it is inherently adaptable to variations and collaborative by virtue of its built-in exploration-exploitation mechanism and swarm nature. Meanwhile, the BI part also preserves historical information, which enables to re-evaluate the optimization results when the environment changes.

\section{Performance Evaluation}
\begin{figure}
\centering
{\includegraphics[width=2.68in]{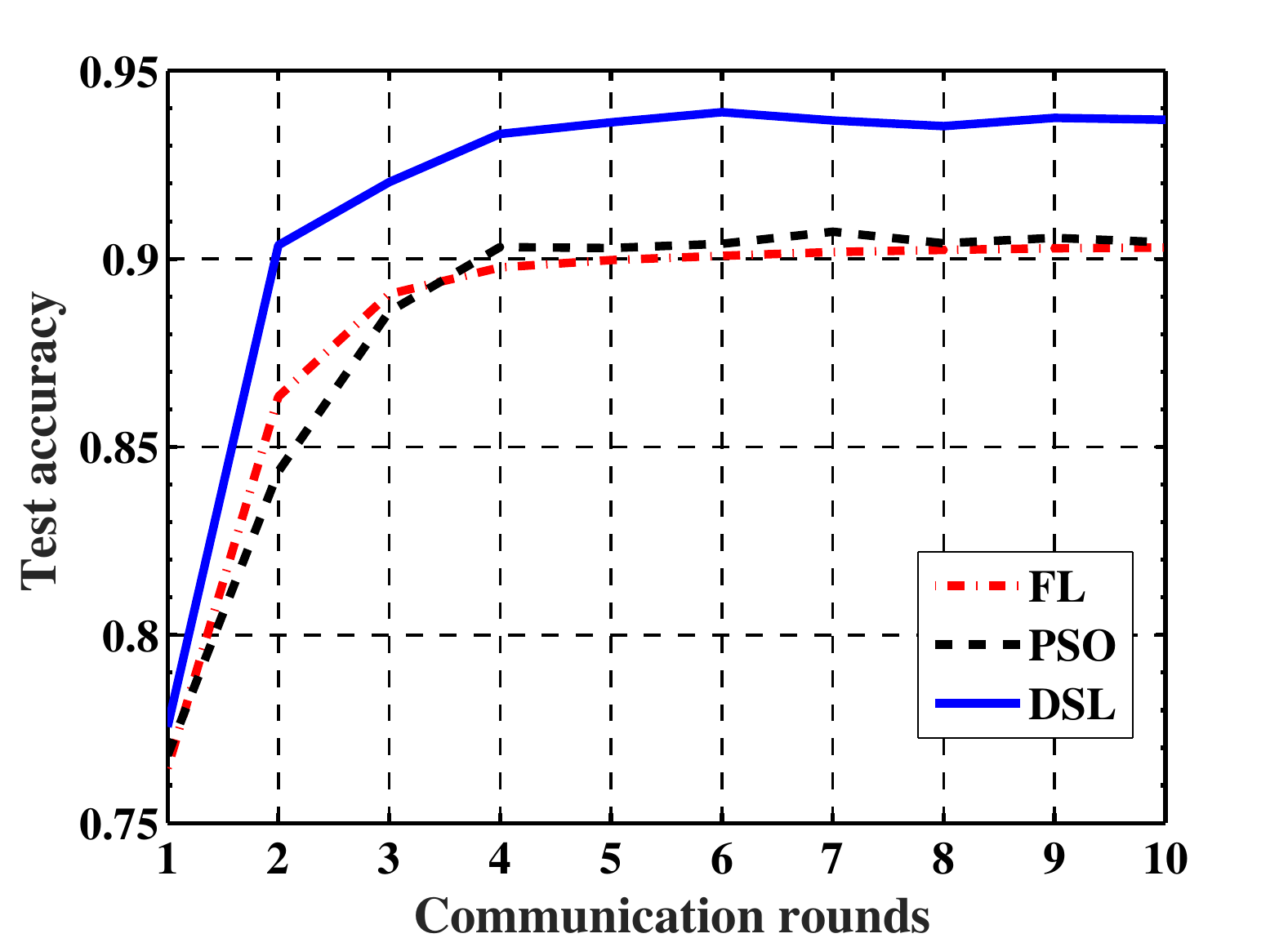}}
 \caption{\small Performance of different methods in i.i.d. cases.}
 \label{fig:iid}
\end{figure}

\begin{figure}
\centering
{\includegraphics[width=2.68in]{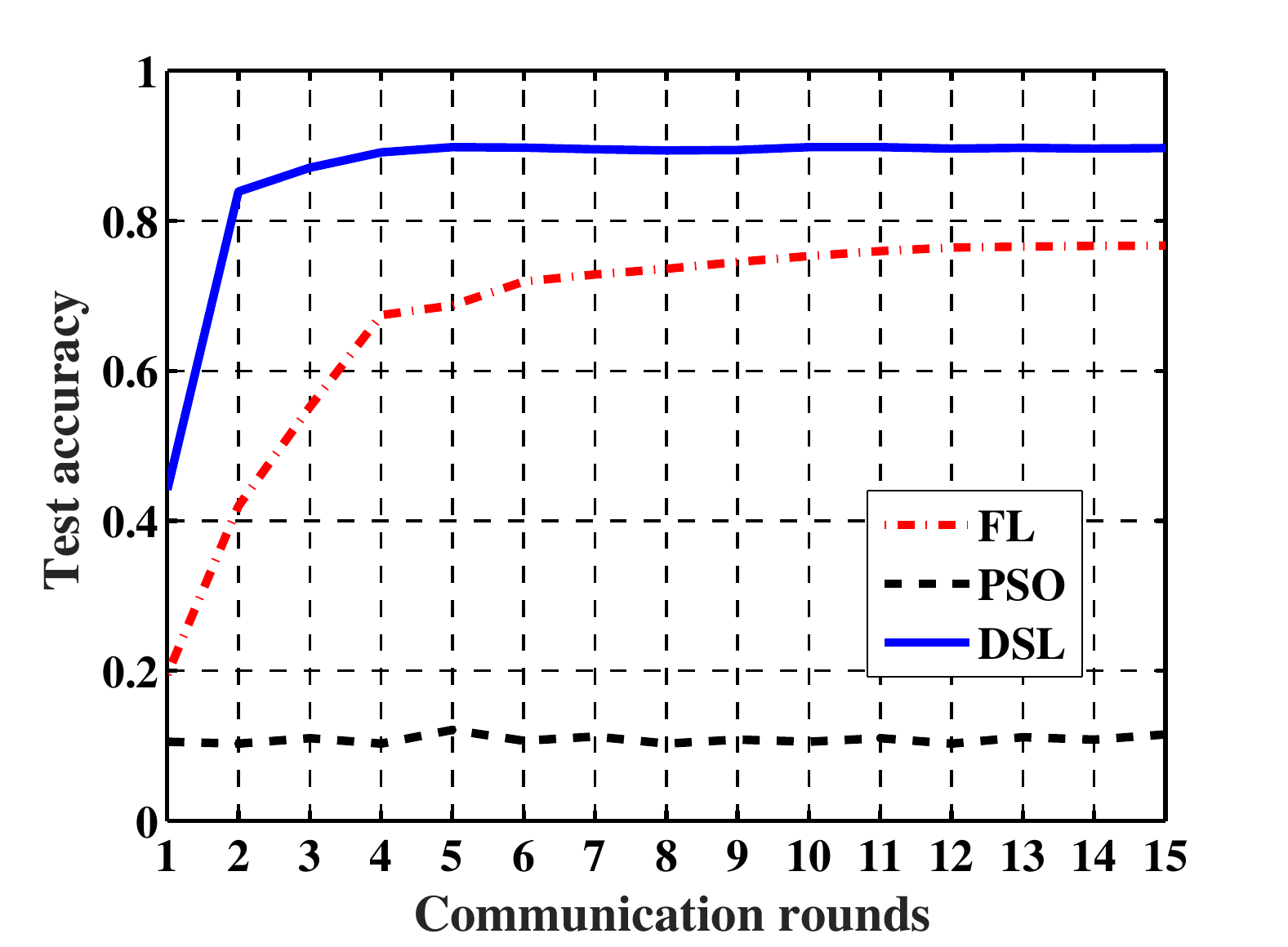}}
 \caption{\small Performance of different methods in non-i.i.d. cases.}
 \label{fig:noniid}
\end{figure}

\begin{figure}
\centering
{\includegraphics[width=2.68in]{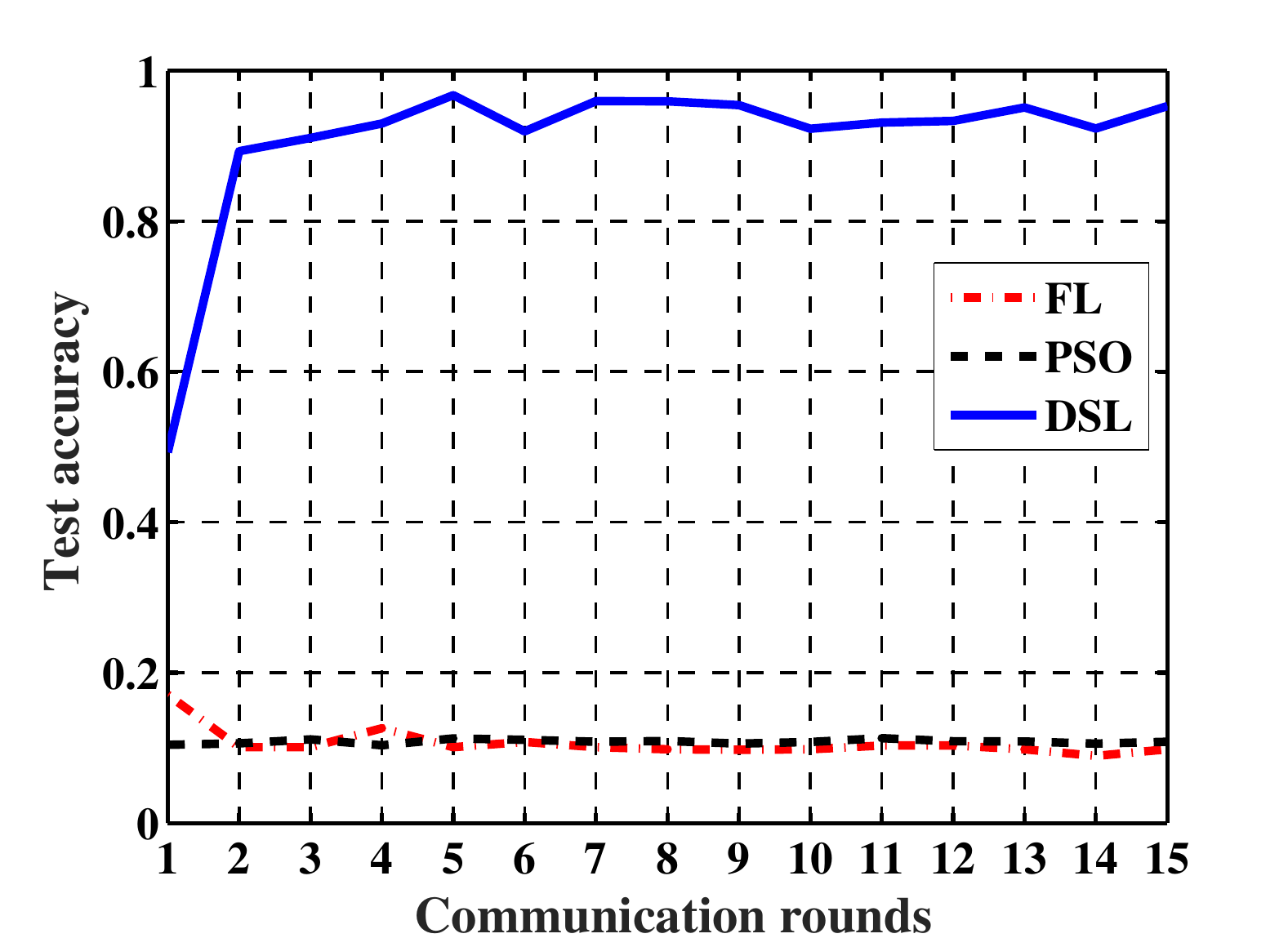}}
 \caption{\small Performance of different methods under 
 Byzantine attacks.}
 \label{fig:attack}
\end{figure}

\noindent
DSL with $50$ distributed workers has been tested on a handwritten-digit classification task using the widely-used MNIST dataset~\cite{JSTSP2022DSL}, compared with FL and PSO as the benchmark methods.
The common learning model is a five-layer Convolutional Neural Network with $44426$ trainable parameters, using a cross-entropy loss function. As shown in Figure~\ref{fig:iid}, DSL outperforms FL and PSO in achieving the highest accuracy.
At the same performance level, the communication round of DSL is less than that of FL or PSO.
Such advantage of DSL becomes more obvious in the non-i.i.d. case in Figure~\ref{fig:noniid}, where PSO fails to work properly because its underlying assumption of a common loss function for all workers can only be approximately true in the i.i.d. case.
Further, DSL can effectively defend Byzantine attacks by proactively screening and removing potential attackers, as Figure~\ref{fig:attack} indicates.
These simulation results verify that the integration of AI and BI leads to evident improvement of learning accuracy, communication efficiency and robustness, which brings significant benefits to edge learning among massive IoT devices with small-volume data.


\section{Conclusion and Future Work}
\noindent
This article studies a new DSL paradigm for edge IoT systems, by bridging federated learning with swarm optimization. With theoretical backing, efficient information extraction and exchanging schemes are designed for high efficiency in communication and computation of model updates. Robust DSL solutions are developed to cope with data heterogeneity, node/link failure and Byzantine attack. Dynamic optimization techniques are also investigated to handle online streaming data. While DSL provides promising improvement over existing methods, some open issues and future directions are summarized below.
\subsubsection{Decentralized Topology}
For DSL adopting a star topology, the central server plays a crucial role in synchronizing the iterative updating algorithms running at distributed workers during parallel computing, and performing centralized communication resource allocation. However, it is more challenging for a fully decentralized network, because each worker has to make autonomous decisions on the global learning task, via communication with adjacent workers only. Nevertheless, due to the hindrance of worker synchronization requirements and considering the decentralized topology of many practical IoT networks, there is a growing interest in decentralized DSL to take advantage of its built-in robustness to node failure and asynchronous computing in heterogeneous environments.
\subsubsection{Security Provisioning}
Research on DSL has just started to seek understanding of system-level design and algorithm-level development, with little effort on security provisioning yet. DSL over-the-air  leaves edge learning systems still vulnerable to malicious attacks, because there is no mechanism to retrieve individual messages that help to identify attackers. While BEV is designed to limit the worst impact on learning performance from attackers \cite{fan2022bev}, it takes a best-effort approach that cannot fully protect DSL from harmful attacks. For proactive security provisioning, it is urgent to utilize cybersecurity techniques in DSL, such as authentication and cryptographic mechanisms, while keeping the overhead and complexity low.
\subsubsection{Efficiency-Robustness Tradeoff}
Due to the natural tradeoff between efficiency and robustness, the adopted robust aggregation measures may affect the convergence behavior of iterative DSL algorithms. For DSL over-the-air, multiple selected workers do not consume extra bandwidth, but do increase the total  transmit energy in the DSL systems. Therefore, the improved robustness via multi-worker diversity comes at the expense of reduced energy efficiency. Moreover, the non-smooth total variation penalty term may slow down the convergence speed of DSL algorithms and cause convergence errors. Understanding of such tradeoff is vital, which sheds light on the impacts of AI learning and BI optimization parameters on DSL convergence behavior.
\subsubsection{Connection to Other BI Techniques}
DSL is still at its infancy, which calls for substantial efforts to leverage versatile BI-inspired swarm optimization techniques for realizing the full potential of DSL in IoT-driven edge learning regimes. Besides PSO-enabled DSL illustrated in this article, there are other notable swarm intelligence techniques that are advantageous in solving various task-oriented optimization problems, including ant colony optimization, artificial bee colony, bacteria foraging algorithms, to name a few. These BI-inspired techniques have not been fully explored in edge learning and IoT applications, which open up opportunities in developing efficient, robust and adaptive enabling techniques by integrating AI with BI of broad scope.
\subsubsection{Prototype Development}
\begin{figure}
\centering
{\includegraphics[width=2.6in]{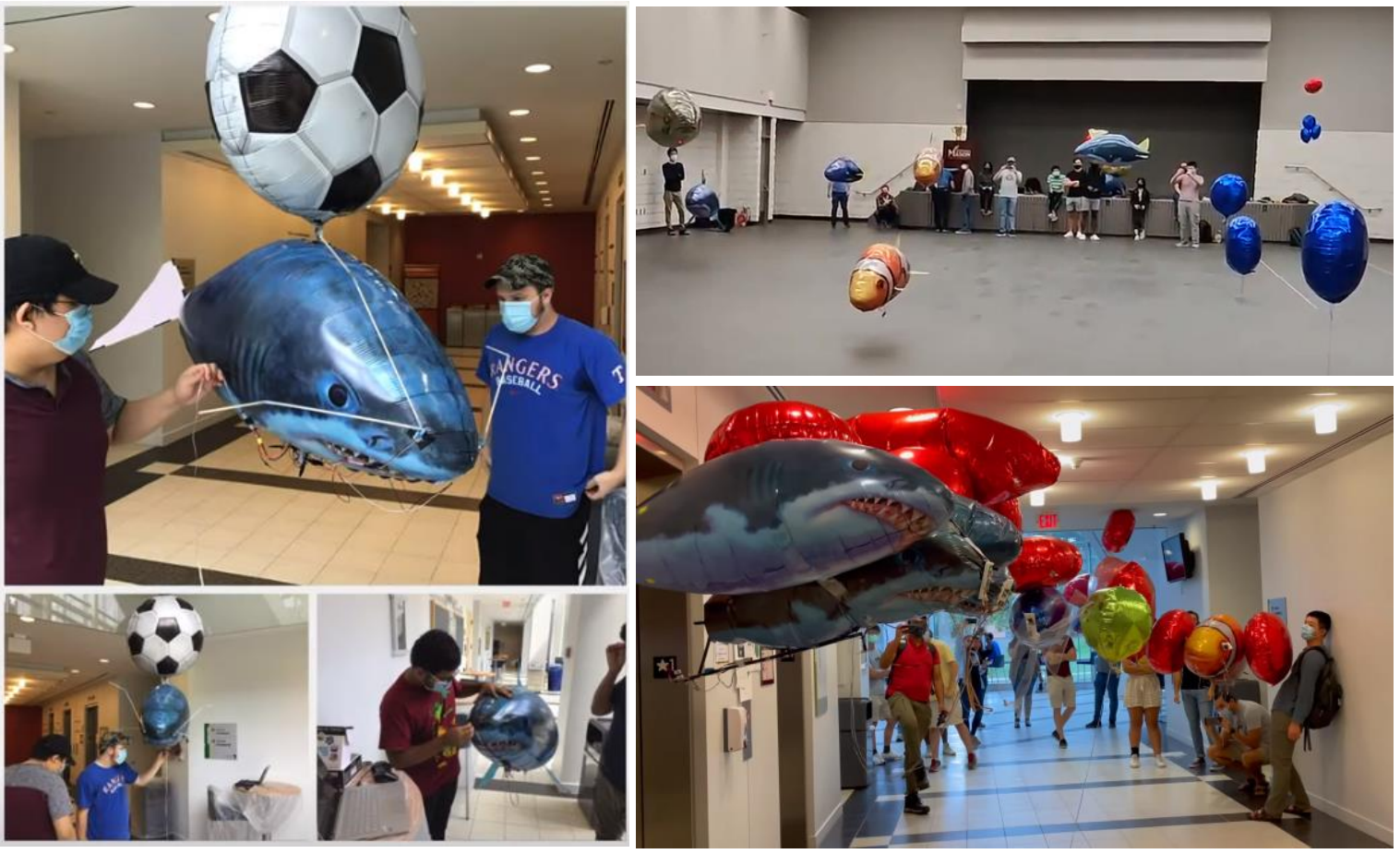}}
 \caption{\small SPARX platform.}
 \label{fig:SPARX}
\end{figure}

DSL is expected to have broad applications in practical distributed systems such as robotic swarms. For validating tests, DSL solutions are planned to be implemented on the {Swarming Platform for Autonomous Robots X (SPARX)} built by the \emph{Patriot Pilots} team at George Mason University. As shown in Figures~\ref{fig:DSL_EdgeIoT}~and~\ref{fig:SPARX}, the SPARX platform consists of a swarm of lighter-than-air balloons as blimp robots (a flying IoT system) where each blimp has limited communication and computation capability. SPARX is designed to play aerial soccer matches in the Defend the Republic Blimp Competition hosted by the Office of Naval Research.
SPARX players will employ the DSL techniques to train learning models for classifying objects in a soccer game, e.g., round ball, rectangular gate, triangular corner flag, and diamond field center.

\section*{Biographies}
\scriptsize{
\noindent Yue Wang [SM'22] (ywang56@gmu.edu) received the Ph.D. degree in communication and information system from Beijing University of Posts and Telecommunications, China, in 2011. Currently, he is a Research Assistant Professor with the Electrical and Computer Engineering Department of George Mason University, Fairfax, VA, USA. Previously, he was a Senior Research Engineer with Huawei Technologies Co., Ltd., China. His general interests lie in the areas of signal processing, wireless communications, artificial intelligence, and their applications in cyber physical systems. His current research focuses on compressed sensing, massive MIMO, mmWave communications, NOMA, cognitive radios, Internet of Things, DoA estimation, high-dimensional data analysis, and distributed optimization and learning.
\vspace{0.08in}

\noindent Zhi Tian [F'13] (ztian1@gmu.edu) has been a professor with the Electrical and Computer Engineering Department of George Mason University since 2015. Prior to that, she was on the faculty of Michigan Technological University from 2000 to 2014. She served as a Program Director at the US National Science Foundation from 2012 to 2014. Her research interest lies in the areas of wireless communications, statistical signal processing, and machine learning. Current research focuses on massive MIMO, millimeter-wave communications, and distributed network optimization and learning. She was an IEEE Distinguished Lecturer for both the IEEE Communications Society and the IEEE Vehicular Technology Society. She served as Associate Editor for IEEE Transactions on Wireless Communications and IEEE Transactions on Signal Processing. She was the Chair of the IEEE Signal Processing Society Big Data Special Interest Group, and a Member-of-Large of the IEEE Signal Processing Society (2019-2021). She received the IEEE Communications Society TCCN Publication Award in 2018.
\vspace{0.08in}

\noindent Xin Fan [M'22] (fanxin@bjtu.edu.cn) received the B.E. degree and M.E. degree from School of Electronic and Information Engineering, Beijing Jiaotong University, Beijing, China, in 2016 and 2018, respectively. He is currently a Ph.D. student in Beijing Jiaotong University from 2018, and a visiting Ph.D. student in the Electrical and Computer Engineering Department of George Mason University, Fairfax, VA, USA, from 2020. His current research interests lie in the areas of wireless communications, security and privacy, optimization, statistical signal processing, blockchain, and machine learning.
\vspace{0.08in}

\noindent Yan Huo [SM'20] (yhuo@bjtu.edu.cn) received the B.E. and Ph.D. degrees in communication and information system from Beijing Jiaotong University, Beijing, China, in 2004 and 2009, respectively. He was a visiting scholar with the Department of Computer Science, The George Washington University, from 2015 to 2016. He is currently a Professor with the School of Electronics and Information Engineering, Beijing Jiaotong University. His current research interests include wireless communications, physical layer security, privacy protection, and edge computing. He has served as an associate editor for the IEEE Access and a Reviewer for a number of journals, including the IEEE Wireless Communications, the IEEE Internet of Things Journal, the IEEE Transactions on Wireless Communications, the IEEE Transactions on Vehicular Technology, and the IEEE Transactions on Mobile Computing.
\vspace{0.08in}

\noindent Cameron Nowzari [M'13] (cnowzari@gmu.edu) received the Ph.D. degree in mechanical engineering from the University of California, San Diego, CA, USA, in 2013. He then held a postdoctoral position with the Electrical and Systems Engineering Department at the University of Pennsylvania, Philadelphia, PA, USA, until 2016. He is currently an Associate Professor with the Electrical and Computer Engineering Department, George Mason University, Fairfax, VA, USA. His current research interests include dynamical systems and control, distributed coordination algorithms, robotics, event- and self-triggered control, Markov processes, network science, spreading processes on networks, and the Internet of Things. He has received several awards, including the American Automatic Control Council's O. Hugo Schuck Best Paper Award, the IEEE Control Systems Magazine Outstanding Paper Award, and the International Conference on Data Mining Best Paper Award.
\vspace{0.08in}

\noindent Kai Zeng [M'05] (kzeng2@gmu.edu) received the Ph.D. degree in electrical and computer engineering from Worcester Polytechnic Institute (WPI), in 2008. He is an associate professor with the Department of Electrical and Computer Engineering, Department of Computer Science, and Center for Secure Information Systems, George Mason University. He was a postdoctoral scholar with the Department of Computer Science, University of California, Davis (UCD) from 2008 to 2011. He worked with the Department of Computer and Information Science, University of Michigan - Dearborn as an Assistant Professor from 2011 to 2014. He was a recipient of the U.S. National Science Foundation Faculty Early Career Development (CAREER) award in 2012. He won Excellence in Postdoctoral Research Award at UCD in 2011 and Sigma Xi Outstanding PhD Dissertation Award at WPI in 2008. He is an editor of the IEEE Transactions on Cognitive Communications and Networking. His current research interests include cyber-physical system security and privacy, physical layer security, network forensics, and cognitive radio networks.
\vspace{0.08in}
}

\end{document}